\documentclass{article}
\usepackage{graphicx} 
\usepackage{booktabs} 
\usepackage{url}
\usepackage{tabularx}
\usepackage[english]{babel}
\usepackage{subcaption}

\usepackage[utf8]{inputenc}
\usepackage{etex}
\usepackage{amscd}
\usepackage{amsfonts}
\usepackage{amssymb}
\usepackage{amsmath}
\usepackage{mathtools}
\usepackage{curves}
\usepackage{eucal}
\usepackage{epsfig}
\usepackage{enumerate}
\usepackage{extarrows}
\usepackage{fullpage}
\usepackage{hyperref}
\usepackage[utf8]{inputenc}
\usepackage{ifthen}
\usepackage{latexsym}
\usepackage[usenames,dvipsnames]{pstricks}
\usepackage{pst-grad} 
\usepackage{pst-plot} 
\usepackage[all]{xy}
\usepackage{tikz}
\usepackage{verbatim}
\usepackage{xargs}

\usepackage{amsmath}

\newcommand\norm[1]

\newcounter{comments}
\usepackage{comment}

\author{Logan Rose$^1$, Jonathan Martinez$^2$, Juho Kim$^3$, Jing Qin$^1$, Boris Aguilar$^3$, David Murrugarra$^1$}

\begin{document}
\title{Personalizing Cancer Models under Data Scarcity via Parameter Decomposition}

\maketitle
{\footnotesize
     \centerline{$^1$Department of Mathematics,
      University of Kentucky, Lexington, KY 40506, USA}
}
{\footnotesize
     \centerline{$^2$Department of Quantitative and Computational Biology, University of Southern California, CA 90089, USA}
}
{\footnotesize
     \centerline{$^3$Institute for Systems Biology,
      Seattle, WA 98109}
}

\begin{abstract}
Personalized cancer modeling for clinical applications requires robust and efficient parameter calibration, particularly in settings with limited patient data. This need is especially critical for medical digital twins (MDTs), which are virtual representations of disease continuously updated using longitudinal patient measurements. In this work, we propose a novel parameter personalization framework for dynamical cancer models under data scarcity. Our approach decomposes selected model parameters into a common component, shared across patients, and a personalized component, which is patient-specific and can be updated as new data become available. The common component captures population-level structure and is estimated once, providing an informed prior that enables rapid and accurate personalization. We demonstrate the effectiveness of this framework using synthetic data generated from canonical dynamical systems, such as logistic growth models with optimized treatment interventions. Our results show that parameter decomposition significantly improves calibration performance in limited-data regimes, facilitating fast and reliable personalization and supporting the development of patient-specific cancer models and MDTs.

\end{abstract}

\section{Introduction}

The development of predictive dynamical models that can be integrated into Medical Digital Twins (MDTs) for clinical applications requires effective strategies to adapt model behavior to the specific characteristics of individual patients. Such personalization is particularly critical for diseases characterized by strong inter‑patient heterogeneity, including acute myeloid leukemia (AML)~\cite{hoffmann2020differential} and other hematological malignancies, where variability in disease mechanisms can confound treatment selection and outcome prediction. However, estimating patient‑specific parameters of dynamical disease models from longitudinal clinical data remains a major challenge, largely due to the nonlinear nature of the underlying biological processes and the limited availability of patient‑specific time‑resolved data.

Model personalization in disease modeling—especially in mathematical oncology—is difficult for two fundamental reasons. First, our incomplete understanding of the biological mechanisms governing disease progression restricts the development of mechanistic models with reliable predictive power. Only a limited number of examples exist in which mathematical models have been successfully incorporated into a clinical practice, such as digital twins for type‑1 and type‑2 diabetes~\cite{kovatchev2025human,wang2025artificial} and cardiovascular disease~\cite{sel2024building,qian2025developing,thangaraj2024cardiovascular}. Moreover, modeling complex diseases often requires a large number of parameters to capture interacting mechanisms across multiple biological scales. These parameters cannot be readily inferred from patient data, motivating the need for modeling approaches that minimize the number of essential mechanisms while retaining clinical relevance.

Second, model parameters are inherently difficult to calibrate using patient‑specific longitudinal data because such data are typically sparse. Clinical measurements relevant to cancer modeling are often collected only during infrequent clinical visits, resulting in short and noisy time series~\cite{hoffmann2020differential}. This setting contrasts sharply with domains where data‑driven and machine‑learning approaches have proven successful due to the availability of large, dense datasets. While high‑dimensional datasets such as genomics or transcriptomics are increasingly available, they generally consist of single snapshots rather than longitudinal trajectories and therefore fail to capture the dynamic nature of disease progression. Without effective methods for parameter personalization under limited longitudinal data, MDTs cannot continuously update and improve as new patient data become available—a defining feature of digital twin technologies.

Model personalization is especially necessary for cancer modeling and their integration into MDTs. Here we use available mathematical models as a proof of concept to generate synthetic data to test our personalization approaches. Existing mathematical models focus on the competitive dynamics between leukemic and healthy stem cell populations but lack systematic approaches for patient‑specific parameterization under data constraints.

In this work, we propose a novel framework for model personalization under limited longitudinal data, based on a principled decomposition of model parameters. By separating parameters according to their roles in governing shared versus patient‑specific dynamics, our approach enables reliable calibration by proving an efficient prior for personalization. Using existing mathematical models as a proof of concept and synthetic data to emulate clinical constraints, we demonstrate that parameter decomposition markedly improves calibration performance in data‑limited regimes. Our results provide a practical pathway toward fast, reliable personalization of cancer models and support the broader development of adaptive, patient‑specific MDTs.

\section{Methods}

\subsection{Proposed Parameter Decomposition}
In this work, we recover dynamical models from longitudinal patient data and subsequently personalize them. Our method of personalization hinges on decomposing model parameters $\mathbf{\theta}_i$ into two components: a \textit{common component} ($\mathbf{\theta}_c$) to represent a commonality between patients, and a \textit{personalized component} ($\mathbf{\theta}_{p_i}$) which represents the deviation from the common component. The common component is estimated using a retrospective patient dataset; this is similar to an offline training process used in machine learning methods. The common component is the same for all patients in the training dataset. The personalized components vary from patient to patient and represent real-time updates made by Medical Digital Twins when given new patient-specific data. The personalized components will change as new data becomes available.

Suppose we are given an MDT with underlying dynamical system 
\begin{equation*}
\frac{d}{dt} \mathbf{x}(t) = \mathbf{f}(\mathbf{x}(t), \mathbf{\theta}, \mathbf{u}(t))
\end{equation*} with state variables
$\mathbf{x}(t) \in \mathbb{R}^n$, input variables $\mathbf{u}(t) \in \mathbb{R}^{\ell}$, vector of model parameters $\mathbf{\theta}$, and differentiable function $f$, where $f: \mathbb{R}^n \times \mathbb{R}^{\ell} \rightarrow \mathbb{R}^n$. Here, $f$ denotes the set of governing equations for a given dynamical system. The input for our models is a set of longitudinal data collected at $k$ time points $t_1, \cdots, t_k$. Suppose there exist observation data $\{\mathbf{X}_i\}_{i=1}^{m}$ for $m$ patients and that the underlying dynamical system has $n$ parameters.

 Let $\mathbf{\theta}_i=\begin{bmatrix} \xi_{i_1} & \ldots& \xi_{i_n} \end{bmatrix}^T$ be the vector of parameters for the $i$th patient. Then  we propose the following decomposition of parameters: \begin{equation}\mathbf{\theta}_i=\mathbf{\theta}_c+\mathbf{\theta}_{p_i} \label{L1} \end{equation} where $\mathbf{\theta}_c=\begin{bmatrix} \xi_{c_1} & \ldots & \xi_{c_n} \end{bmatrix}^T$ and $\mathbf{\theta}_{p_i}=\begin{bmatrix} \xi_{p_{i_1}} & \ldots & \xi_{p_{i_n}} \end{bmatrix}^T$. Note $\mathbf{\theta}_c$ represents the vector of common components and $\mathbf{\theta}_{p_i}$, the vector of personalized components for the $i$th patient. 
 Here, $\mathbf{\theta}_c$ is the same for all patients in a population and is estimated once. Conversely, $\mathbf{\theta}_{p_i}$ represents the personalized component and differs from patient to patient; it is the deviation from the common component. There are two main goals of our decomposition: to recover a model which fits even limited data well and to minimize the personalized components.

 \subsection{Mathematical Formulation of the Approach}

The common and personalized components ($\mathbf{\theta}_c$ and $\mathbf{\theta}_{p_i}$ respectively) are recovered in two phases. First, $\mathbf{\theta}_c$ is estimated in an offline training phase using a retrospective patient dataset. The second phase represents a real time update; the personalized component $(\mathbf{\theta}_{p_i})$ is updated as new patient measurement data come in, while $\mathbf{\theta}_c$ is fixed.
The utility of the decomposition ($\mathbf{\theta}_i = \mathbf{\theta}_c+\mathbf{\theta}_{p_i}$) is that when given data $\mathbf{X}_d$ for a new patient, $d$, we can more quickly recover the corresponding parameters for that patient $\theta_d=\theta_c+\theta_{p_i}$ using the previously estimated $\mathbf{\theta}_c$ as a starting point.

The input for our training phase is a set of observations for $m$ patients ($\{ \mathbf{X_i}\}_{i=1}^{m}$). We will recover the common ($\mathbf{\theta}_c$) and personalized components ($\mathbf{\theta}_{p_i})$ by constructing and minimizing a loss function. This loss function contains two terms for each patient $\mathbf{X_i}$: an error term, $L_{1_i}$, and a regularizer, $L_{2_i}$. The error term, $L_{1_i}$, is used to minimize the error in estimating the components, while the regularizer, $L_{2_i}$ is used to minimize the personalized components $(\mathbf{\theta}_{p_i})$ to avoid overfitting to the training set. The terms of our loss, $L_{1_i}$ and $L_{2_i}$ for each trajectory $\mathbf{X}_i$ are given below:

\begin{equation*}
\begin{aligned}
L_{1_i} &= \|\mathbf{\dot{X}}_i - \mathbf{\hat{\dot{X}}}_i\|_2^2 
\quad\quad
L_{2_i} = \lambda \|\mathbf{\theta}_{p_i}\|_2^2.
\end{aligned}
\end{equation*} 
where $\mathbf{\dot{X}}_i$ is the derivative of $\mathbf{{X}}_i$ (computed by finite difference) and  $\hat{\dot{\mathbf{X}}}_i=\Theta(\mathbf{X})(\mathbf{\theta}_c+\mathbf{\theta}_{p_i})$. Here, we assume the data ($\{ \mathbf{X_i}\}_{i=1}^{m}$) are represented by a dynamical system, $f$. Additionally, $\Theta(\mathbf{X})$ is a library of basis functions which may appear in $f$ and $(\mathbf{\theta}_i=\mathbf{\theta}_c+\mathbf{\theta}_{p_i})$ are the parameter values estimated using the common and personalized components. Thus, $\Theta(\mathbf{X})(\mathbf{\theta}_c+\mathbf{\theta}_{p_i})$, represents the derivative approximated by dynamical system $f$ with parameters $\theta_i=\theta_c+\theta_{p_i}$.  Note that $\lambda$ in the $L_{2_i}$ term is a sparsity coefficient used to keep the personalized component small. 
Note that $L_{1_i}$ is our error term; it represents the distance between the true derivative $\dot{\mathbf{X}}_i$ and an estimated derivative $\hat{\dot{\mathbf{X}}}_i$. Additionally, $L_{2_i}$ is a regularizer used to keep the personalized component small. 

Next, we define the total loss function as follows:
\[L(\theta)=\sum_{i=1}^{m}(L_{1_i}+L_{2_i})=\sum_{i=1}^{m}(||\mathbf{\dot{X}_i}-\mathbf{\hat{\dot{X}}}_i||_2^2+||\mathbf{\theta}_{p_i}||_2^2).\]
Then, to recover the common/personalized components, we solve:
\begin{equation*}
\min_{\mathbf{\theta}_c,\mathbf{\mathbf{\theta}_p}} L(\mathbf{\theta}).
\end{equation*}
There are many methods to solve this problem; below, we present one such approach.

\subsection{Solving the Loss Function}

We solve for $\mathbf{\theta}_c$ and $\mathbf{\theta}_{p_i}$ in two steps. First, we modify our loss function slightly and recover the value of $\mathbf{\theta}_c$; then we fix $\mathbf{\theta}_c$ to be that found in the first step and solve our original loss function ($L(\theta)$) for $\mathbf{\theta}_{p_i}$. In both steps, the Adam algorithm \cite{kingma2015adam} is used to solve for the components.

\textbf{Step 1}: Find $\mathbf{\theta}_c$ using Adam. Suppose we have $m$ parameters and $n$ patients. Let $\mathbf{\theta}=\begin{bmatrix} \mathbf{\theta_1}& \ldots &\mathbf{\theta_m} \end{bmatrix}^T$ be a matrix of parameters  where each $\mathbf{\theta}_i=[\mathbf{\theta}_{i_1}\cdots \mathbf{\theta}_{i_n}]^T$, the set of all parameters for the $i$th patient. In this case, $\mathbf{\theta}$ is a $m \times n$ matrix where the $i$th row ($1 \leq i \leq m$) contains all parameter values of the $i$th patient, and the $j$th column $(1 \leq j \leq m)$ contains the values for the $j$th parameter across all patients. Then, we recover each $\theta_i$ by solving the following loss function:

\begin{equation*}
 \min_{\mathbf{\theta}}
 L(\mathbf{\theta})=\sum_{i=1}^{m} \| \dot{\mathbf{X_i}}-\Theta(\mathbf{X_i})\mathbf{\theta}_i\|_2^2.
 \end{equation*}

 Here we do not decompose the parameters as $\mathbf{\theta}_i=\mathbf{\theta}_c+\mathbf{\theta}_{p_i}$; we merely estimate the parameter values $\mathbf{\theta}_i$ for each patient. Then the common component is given as follows:
$\mathbf{\theta}_c=\begin{bmatrix} \mathbf{\theta}_{c_1}& \ldots &\mathbf{\theta}_{c_n} \end{bmatrix}^T$. Each $c_j$ is the common component for the $j$th parameter. We assume that the values recovered for the $j$th parameter have probability distribution $\pi_j$. Then $\mathbf{\xi}_{c_i}=\mu_j$, where $\mu_j$ is the mean of $\pi_j$.
\newline
 
\textbf{Step 2}: We recover the personalized components $(\mathbf{\theta_{p_i}})$. First, we fix the common component, $\mathbf{\theta}_c$, to be that recovered in Step 1. Then, using the Adam optimizer, we minimize the original loss function $L(\theta)$ with respect to $\mathbf{\theta}_{p_i}$, i.e.:
\begin{equation*}
 \min_{\mathbf{\theta_p}}\sum_{i=1}^{m}
||\mathbf{\dot{X_i}}-\Theta(\mathbf{X_i})(\mathbf{\theta}_c+\mathbf{\theta}_{p_i}) ||_2^2+\lambda||\mathbf{\theta}_{p_i}||_2^2.
\end{equation*}

\subsection{Assessing the Method}

We will assess the accuracy of our method using synthetic data generated from logistic growth models with optimized treatment interventions. We do so in two phases: training and adaptation.

\subsubsection{Training Phase}

First, using a known dynamical system, $\frac{d\mathbf{x}}{dt}=f(\mathbf{x},\mathbf{\theta},t)$, we generate $m$ trajectories $\{ \mathbf{X}_i\}_{i=1}^{m}$ in Python by integrating $f$. Each trajectory $\mathbf{X}_i$, represents the state data for the $i$th patient. Using the method outlined in the previous section, we first recover the common component $\theta_c$ and then the personalized component for each trajectory $\theta_{p_i}$. Then, for each $\mathbf{X}_i$, an estimated trajectory $\hat{\mathbf{X}}_i$ with parameters $\mathbf{\theta}_i=\mathbf{\theta}_c+\mathbf{\theta}_{p_i}$ is generated by integrating $f$ (the same dynamical system used to generate $X_i$). To assess the accuracy of the method, the Mean-Squared Error (MSE) of true trajectory $\mathbf{X}_i$ and estimated trajectory $\hat{\mathbf{X}}_i$ are computed. Here, 
\[MSE=\frac{1}{n} \|\mathbf{X}-\hat{\mathbf{X}}\|_2^2.\] 
The average MSE for all trajectories is then computed.

\subsubsection{Adaptation Phase} 

We fix $\mathbf{\theta}_c$ to be that computed during the training phase. Then, $\ell$ new trajectories, $\{ \mathbf{Y_j}\}_{j=1}^{\ell}$, which represent a set of new patients, are generated. The personalized components, $\{\mathbf{\theta_{p_j}}\}_{i=1}^{j}$, are recovered for each trajectory $\mathbf{Y_j}$. As in the training phase, for each $Y_j$, an estimated trajectory, $\hat{Y_j}$, is generated by integrating $f$. The average MSE values $\mathbf{Y_j}$ and $\hat{Y_j}$ are then computed.

\section{Results}

We assess our methods using synthetic data (trajectories) generated by integrating known dynamical models with fixed parameters. Given a trajectory $X_i$, with true dynamical model, $f$ and parameters $\mathbf{\theta}_i$, we will recover an estimated dynamical model $\hat f$ with estimated parameters $\mathbf{\hat{\theta}}_i=\mathbf{\theta}_c+\mathbf{\theta_{p_i}}$. We do so using the two-step method listed above. 
Here, we work with a modified logistic model based on the Hoffmann Cancer model \cite{hoffmann2020differential}.

\subsection{Personalizing Known Dynamical models}

\subsubsection{Reduced model for cancer treatment (Logistic growth)}

We consider a reduced version of the Hoffmann model, which we refer to as the \textit{Logistic Growth with Control} model. This model measures the change in the number of active leukemic cells over time and is given by:

\begin{equation*}
\frac{d\mathbf{y}}{dt}=(p_l-d_k)\mathbf{y}-\frac{p_l}{K_A}\mathbf{y}^2-c\text{*}\mathbf{u(t)}\text{*}\mathbf{y},
\end{equation*}
where $\mathbf{y}$ is the number of active leukemic cells present at time $\mathbf{t}$. This model has four parameters: $p_l$ gives the proliferation rate, $d_k$ the natural death rate, $K_A$ the carrying capacity, and $c$ the death rate of leukemic cells due to chemotherapy. This equation has an input, $\mathbf{u}$, where
\[ \mathbf{u}(t) = \begin{cases} 
          1 & \text{if chemo is applied at time }$t$ \\
          0 & \text{if chemo is inactive at time }$t$ 
       \end{cases}
    .\]

\begin{figure}[h!]
    \centering
    \begin{subfigure}{0.45\textwidth}
        \centering
        \includegraphics[width=\linewidth]{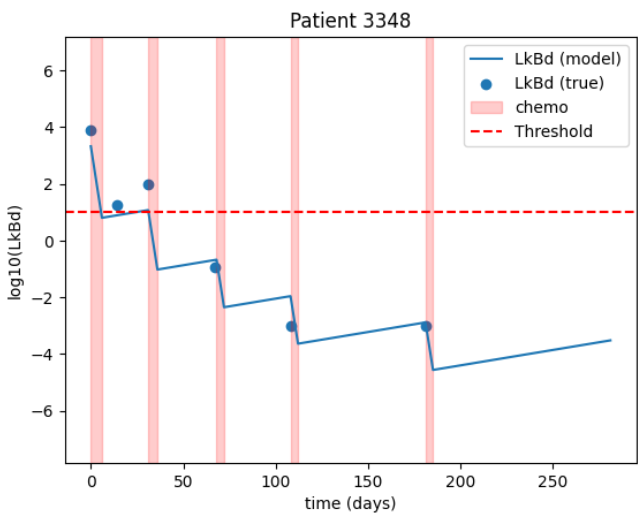}
        
    \end{subfigure}
    \hfill
    \begin{subfigure}{0.45\textwidth}
        \centering
        \includegraphics[width=\linewidth]{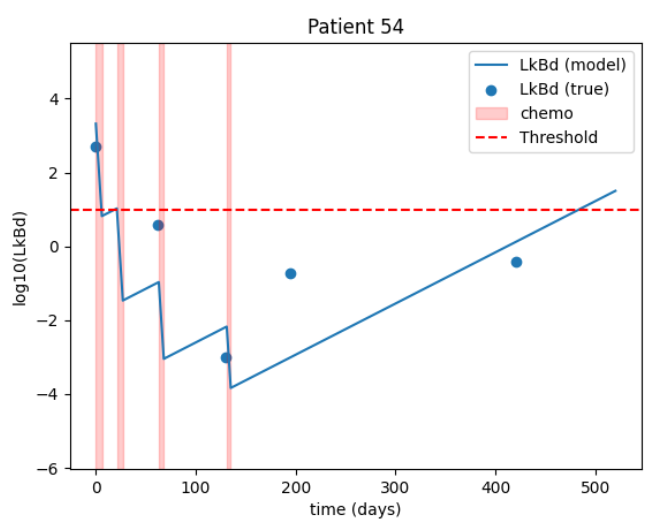}
        
    \end{subfigure}
    \caption{Example of trajectories generated using the logistic growth with control model. Application of chemotherapy treatment is indicated by red stripes, while true observations are denoted by solid dots. Left: Note that the $y$-value falls below a fixed threshold (dotted red line) and stays below. This patient is considered non-relapsed (NR). Right: Note that the $y$-value falls below a fixed threshold (dotted red line) and rises back above. This patient is considered relapsed (R). }
    \label{fig:lgplot}
\end{figure}

Example trajectories of this model are given in Figure~\ref{fig:lgplot}. Note that when $\mathbf{u}=0$, this model will behave like a standard logistic-growth model.

For this model, the parameter $\mathbf{c}$ (death rate due to chemotherapy) is recovered and personalized. This parameter is chosen for personalization because the effects of chemotherapy may vary greatly from patient to patient. In this case, we assume the model is known and all other model parameters are fixed.

 We consider a data set of 275 AML (Acute Myeloid Leukemia) patients which was originally used in \cite{hoffmann2020differential}. This data was pre-fitted to the logistic growth with control model to compute optimal parameter values for each patient. We define a patient as \textbf{Relapsed (R)}, if the $y$ value falls below a fixed threshold $(1$ on a $\log_{10}$ scale) and eventually rises back above the threshold. Conversely, a patient is \textbf{Non-Relapsed (NR)} if the the $y$-value falls below $1$ and does not recover. The original data set was filtered to include only patients with a minimum of 3 NPM1 (leukemic cell counts) observations taken between the beginning of the first and the end of the last treatment cycle. After filtration, there are a total of 171 patients, 47 of which are considered relapsed (R) and 124 that are non-relapsed (NR). 
These experiments were performed on two subsets of the patient data: a set of R patients and a set of NR patients. In each case, $70\%$ of patients were allocated to training and $30\%$ to adaptation. Cross-validation was performed by splitting the data into training and testing over $n=20$ trials.

Trajectories for each patient were generated by fitting $\mathbf{c}$ for each patient beforehand and setting $\mathbf{c}$ equal to the fitted value, fixing $d_k=1/30$, $p_l=0.03$, $K_A=10000$ for all patients, and integrating the  logistic growth with control model numerically in Python. Trajectories were generated from time $t_0=0$ to $t_{final}=500$. The common ($\mathbf{c}_c$) and personalized components ($\mathbf{c}_{p_i}$) of $c$ were recovered for each trajectory. The estimated $\mathbf{c}$ value for each trajectory is then given by $\hat{\mathbf{c}}_i=\mathbf{c}_c+\mathbf{c}_{p_i}$. Then for each trajectory, $\mathbf{X}_i$, in the test set, we integrate the logistic growth  model with $\hat{\mathbf{c}}_i=\mathbf{c}_c+\mathbf{c}_{p_i}$ and all other parameters fixed as before. This generates an estimated trajectory, $\hat{\mathbf{X}}$. Then, the average mean squared error (MSE), average relative error (RE), and median and average parameter difference (PD) of $\mathbf{X}$ and $\mathbf{\hat{X}}$ were computed over all trials, where \begin{equation*}
\begin{aligned}
MSE = \frac{1}{n} \|\mathbf{X}-\hat{\mathbf{X}}\|_2^2, \quad
PD = |\mathbf{c_i}-\hat{\mathbf{c}}_i|, \quad
RE = \frac{\|\mathbf{X}-\hat{\mathbf{X}}\|_2}{\|\mathbf{X}\|}. 
\end{aligned}
\end{equation*}

\begin{figure}[h!]
    \centering

    \begin{subfigure}{0.48\textwidth}
        \centering
        \includegraphics[width=\linewidth]{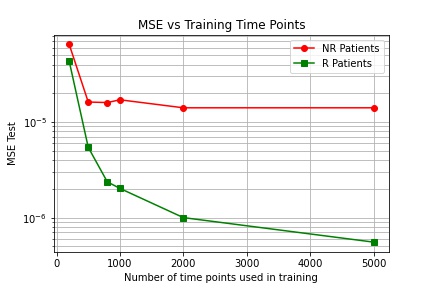}
        \caption{MSE vs. training time}
        \label{fig:MSEoriginal}
    \end{subfigure}
    \hfill
    \begin{subfigure}{0.48\textwidth}
        \centering
        \includegraphics[width=\linewidth]{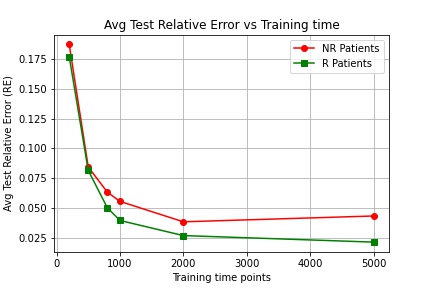}
        \caption{RE vs. training time}
    \end{subfigure}

    \vspace{0.5cm}

    \begin{subfigure}{0.48\textwidth}
        \centering
        \includegraphics[width=\linewidth]{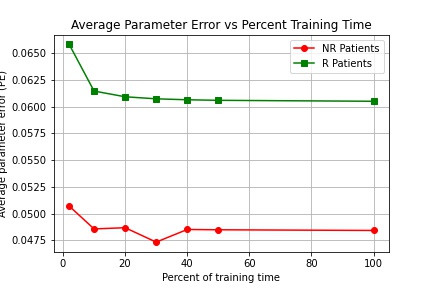}
        \caption{Average parameter difference}
    \end{subfigure}
    \hfill
    \begin{subfigure}{0.48\textwidth}
        \centering
        \includegraphics[width=\linewidth]{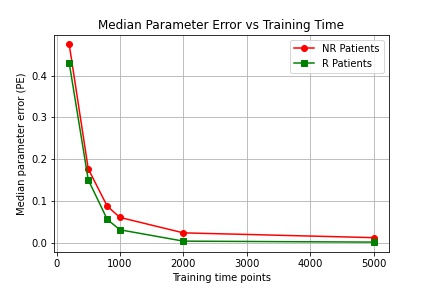}
        \caption{Median parameter difference}
        \label{fig:median_pd_original}
    \end{subfigure}
    
    \caption{Model performance metrics vs. training time for the Logistic Growth model. }
    \label{fig:lgplot2}
\end{figure}

Our results are given by the plots in Figure \ref{fig:lgplot2} . Note that the method estimates each trajectory with minimal MSE. Here, the initial value, $y_0$, is treated as a parameter and fitted beforehand along with the other parameters. These experiments were carried out on 20 trials (20 different training and test sets). Here, training trajectories  are simulated over a time interval of $t_0=0$ to $t_{final}=500$ with $t=200,500,800,1000,2000,5000$ time points. The test trajectories are simulated over a time interval of $t_0=0$ to $t_{final}=500$ with $t=1000$ time points. For all cases, the error values are comparable between R and NR patients, with the relapsed patients having lower error values overall. 

\subsubsection{Logistic Growth with Limited Test Data}

In practice, it is desirable to make predictions quickly after the onset of disease; thus our method must be able to adapt to new trajectories with limited data. 
Here, we fix the training time interval and generate partial test trajectories over a small fraction of the training interval. For example, for a training interval from $t_0=0$ to $t_{final}=10$, we may generate a test trajectory from $t_0=0$ to $t_{final}=1$ (representing $10 \%$ of the training time). Note that the spacing of time points is kept consistent between the training and test trajectories. Using partial test trajectories assesses the ability of the method to quickly learn components from limited information.  

Training trajectories are generated over a time interval from $t_0=0$ to $t_{end}=500$ with $1000$ time points. However, here, we assume test trajectories are generated over a set of limited time intervals ($t_0=0$,$t_{end}=10,50,100,150,200,250,500)$. The spatial arrangement of time points is consistent between the training and the test trajectories. We obtain relatively low error values across all cases, even over the partial time intervals. As before, the initial values for each patient were fitted beforehand. Again, these experiments were carried across 20 trials (20 different training and test sets). Again, we compute the average mean squared error (MSE), relative error (RE), and average and median parameter error (PE) over all trajectories and trials.

Our results are given below in Figure \ref{fig:lgplot3}

\begin{figure}[h!]
    \centering

    \begin{subfigure}{0.48\textwidth}
        \centering
        \includegraphics[width=\linewidth]{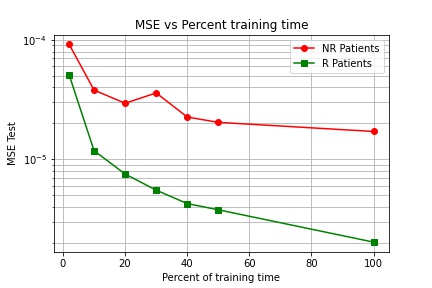}
        \caption{MSE vs. training time}
       
    \end{subfigure}
    \hfill
    \begin{subfigure}{0.48\textwidth}
        \centering
        \includegraphics[width=\linewidth]{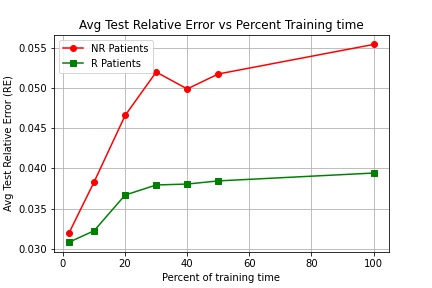}
        \caption{RE vs. training time}
    \end{subfigure}

    \vspace{0.5cm}

    \begin{subfigure}{0.48\textwidth}
        \centering
        \includegraphics[width=\linewidth]{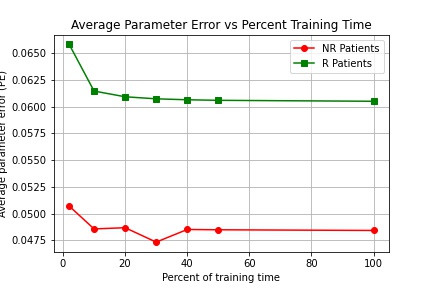}
        \caption{Average parameter error}
    \end{subfigure}
    \hfill
    \begin{subfigure}{0.48\textwidth}
        \centering
        \includegraphics[width=\linewidth]{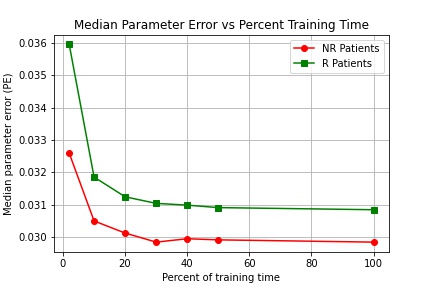}
        \caption{Median parameter error}
        
    \end{subfigure}

    \caption{Model performance metrics vs. percentage training time for the reduced trajectories of the Logistic Growth model. }
    \label{fig:lgplot3}
\end{figure}

MSE Values for all cases remain low, even when only partial test trajectories are provided. This illustrates that our common component can effectively estimate parameters with limited data. A limitation of this experiment is that we assume time points are evenly spaced, while this is often not the case with medical data. We also assume that testing trajectories contain at least 10 days of data; in reality, patient observations may be much more sporadic, with very few observations taken over a long period of time.

\section{Discussion}

We emphasize that the proposed parameter decomposition, $\mathbf{\theta}=\mathbf{\theta}_c+\mathbf{\theta}_{p_i}$,
is only one of several possible ways to separate shared and patient-specific structure. While an additive formulation offers simplicity and interpretability, alternative decompositions—such as weighted sums, multiplicative forms, or low-dimensional manifold representations—may be more appropriate for certain models or biological mechanisms. Exploring such alternatives could further enhance flexibility and capture nonlinear interactions between population-level and individual-specific effects.

Moreover, our current strategy for estimating the common component relies on averaging across individuals, which implicitly assumes approximately normal parameter distributions. This assumption may be restrictive, particularly in heterogeneous patient populations or when parameters exhibit skewness, multimodality, or heavy tails. Future work will therefore focus on more general estimation procedures that relax distributional assumptions, for example by leveraging robust statistics, Bayesian hierarchical modeling, or nonparametric approaches. These extensions would broaden the applicability of the framework and improve its robustness in realistic clinical settings.

\section{Conclusions}

In this work, we introduced a parameter personalization framework designed to address one of the central challenges in patient-specific cancer modeling: reliable calibration under severe data scarcity. By decomposing model parameters into a shared, population-level component and a patient-specific component, our approach leverages common structure across patients while retaining the flexibility needed for individualized predictions. This separation enables the estimation of an informed prior from limited cohort data and supports rapid, stable personalization as new longitudinal measurements become available—an essential requirement for medical digital twins.

Through experiments on synthetic data generated from canonical dynamical cancer models, including logistic growth dynamics with optimized treatment strategies, we demonstrated that parameter decomposition consistently improves calibration accuracy in limited-data regimes. Compared to fully independent personalization, the proposed framework yields faster convergence and more robust estimates, highlighting its potential for real-time or near-real-time clinical use. Beyond improved numerical performance, the framework offers a conceptually simple and extensible strategy for integrating population knowledge into individualized models. Future work will focus on validating the approach with real patient data, extending it to higher-dimensional and mechanistic tumor–immune or treatment-response models, and embedding it within adaptive MDT pipelines to support longitudinal decision-making in oncology.

\section{Acknowledgments}

D.M. was partially supported by the National Science Foundation (NSF Grant No. 2424633) and by a Collaboration Grant (No. 850896) from the Simons Foundation.

\section{Competing interests}

The authors declare no competing financial interests.

\bibliographystyle{elsarticle-num}
\bibliography{references_DT}

\end{document}